\documentclass[journal=aelccp,manuscript=article]{achemso}
\setkeys{acs}{articletitle = true}

\usepackage{graphicx}
\usepackage{dcolumn}
\usepackage{bm}
\usepackage[mathlines]{lineno}

\usepackage{achemso}
\usepackage{amsmath}
\usepackage{color,soul}
\soulregister\cite7
\soulregister\ref7

\usepackage{tabu}
\usepackage{nth}

\usepackage[dvipsnames]{xcolor}
\definecolor{orange}{rgb}{1.0,0.4,0.0}
\usepackage{MnSymbol}
\usepackage{framed}

\newcommand{\textapprox}{{\raise.17ex\hbox{$\scriptstyle\mathtt{\sim}$}}}

\title{Carrier Transport in 2D Hybrid Organic-Inorganic Perovskites: the role of spacer molecules}

\author{Caihong Zheng}
 \affiliation{School of Physical Science and Technology, ShanghaiTech University, Shanghai 201210, China.}

\author{Fan Zheng}
 \affiliation{School of Physical Science and Technology, ShanghaiTech University, Shanghai 201210, China}
 \email{zhengfan@shanghaitech.edu.cn}

\begin{document}
\maketitle

\begin{abstract}
Two-dimensional organic-inorganic hybrid perovskites (2D HOIPs) have been widely used for various optoelectronics owing to the excellent photoelectric properties.
Recently, a great deal of studies have focused on engineering the organic spacer cation in 2D HOIPs to enhance the carrier transport and improve the performance of devices.
However, the selection of organic spacer cations is mostly qualitative without a quantitative guidance.
Meanwhile, the fundamental mechanism of the carrier transport across the organic spacer layer is still unclear.
Here, by using the first-principle non-adiabatic molecular dynamics (NAMD) method, we have studied the transport process of excited carriers between 2D HOIPs separated by the spacer cation layer.
Various types of spacer cations of 2D HOIPs are investigated, where the carrier transport processes are simulated in real-time at atomic levels.
We find that the excited electrons and holes can transfer from single-inorganic-layer 2D HOIP to bi-inorganic-layer 2D HOIP on a sub-picosecond scale, and different types of spacer cations can influence the carrier transfer rate significantly.
Meanwhile, Dion-Jacobson (DJ) phase 2D HOIP leads to a more conductive carrier transport compared to the Ruddlesden-Popper (RP) phase, which is related to the different electron-phonon coupling strengths of these two phases. 
Moreover, we have developed a new method to capture the electron-hole interaction in the frame of NAMD.
This work provides a promising direction to design new materials towards high performance optoelectronics.
\end{abstract}

\newpage

Hybrid organic-inorganic perovskites (HOIPs) have shown excellent photoelectric properties when used as optoelectronic devices (e.g. light-emitting diodes, LED),\cite{TsaiHsinhan2018SLDU,GuoZhenyu2021PETv,ZhangDezhong2023CPfH,KongLingmei2023ASCA} such as high photo-luminescence quantum yield (PLYQ), narrow full width at half-maximum (FWHM), and high spectral adjustability.\cite{YanYinzhou2021SRAi,SadhanalaAditya2015BCTS,WuChen2019ATTL,GuoZhenyu2023HPDi,Chen2021} 
By introducing excess organic amine ions into HOIPs, 2D HOIPs can be formed, which significantly improve the structural stability and reduce the environmental hazards.\cite{MitziD.B1994Cthw,ZhangXiaoli2017HPLD}
HOIPs consist of a series of alternating inorganic perovskite layers and organic spacer layers.\cite{KimYoung-Hoon2015LDMO,QuanLiNa2016LRP}
The large dielectric constant differences between the inorganic perovskite layers and the organic spacer layers of various HOIPs result in the formation of excitons,\cite{doi:10.1021/jacs.8b10851,KatanClaudine2019QaDC} which enhance the radiative recombination rate preferred in an LED device.
In the meantime, the quantum wells (QWs) are naturally formed in the HOIPs enabling the ultra-fast carrier transport.\cite{MitziD.B1994Cthw,WangYa-Kun2020CcoC}
Moreover, by engineering the added organic spacer cations, 2D HOIPs with different dimensions, phases, and properties can be obtained.\cite{FuJianhui2021ESMb,ShiEnzheng2020Thpl}
With these excellent properties, 2D HOIPs have attracted extensive studies in the past few years, and achieved a rapid growth of the external quantum efficiency (EQE) from 1\% to over 20\%.\cite{YuanMingjian2016Peff,LuMin2018SSDa,WatanabeSatoru2020OaSo,SunShuang‐Qiao2023HEHP}
Two types of phases of 2D HOIPs, Ruddlesden-Popper (RP) and Dion-Jacobson (DJ), can be obtained by engineering the composition of organic spacer cations.\cite{doi:10.1021/acsami.0c09651}
The RP phase and DJ phase 2D HOIPs have the chemical formulas $\rm{A^{'}_{2}A_{n-1}B_{n}X_{3n+1} \ and \ A^{''}A_{n-1}B_{n}X_{3n+1}}$, respectively, where $\rm{A^{'}}$ and $\rm{A^{''}}$ are positive monovalent or bivalent cations such as n-butylammonium $\rm{(BA^{+})}$ and p-phenylenediamine $\rm{(PDMA^{2+}})$, A is a small monovalant cation such as $\rm{Cs^{+}}$ and methylammonium $\rm{(MA^{+})}$, B is a metal cation such as $\rm{Pb^{2+}}$ and $\rm{Sn^{2+}}$, and X is a halogen ion such as $\rm{I^{-}}$, $\rm{Br^{-}}$, and $\rm{Cl^{-}}$. 
The value \textit{n} is defined as the number of $\rm{[BX_{4}]^{2-}}$ inorganic perovskite layers, which determines the quantum well width.
Both experiments and calculations have shown that the 2D HOIPs with a large \textit{n} tend to have a small band gap.\cite{ZhaoChunyi2020TLCT,AcharyyaParibesh2019Spso}
However, in the experiment, different \textit{n} layers are usually synthesized simultaneously since the formation energies of different \textit{n} layers are similar.\cite{TakezoeHideo2010Lefo}
Thus, carriers need to transport from a small \textit{n} layer to a large \textit{n} layer.
For this reason, the carriers transportability between different \textit{n} inorganic layers will be critical to the radiation recombination efficiency and PLQY in an LED device.\cite{WangNana2016Pldb}

The organic spacer layer with a large band gap forms a barrier that hinders the carrier transport.\cite{https://doi.org/10.1002/adma.202005570} 
Moreover, the weak van der Waals (vdW) interaction between the organic spacer layers connecting two RP phase 2D HOIPs further suppresses the carrier transport. 
\cite{LiHan2020LPUP}
To enhance the carrier transport across the organic spacer layer,
for instance, Zheng et al found that the shorter aliphatic spacer cation facilitate the energy transfer, which can compete with the fast high-order carrier recombination and consequently improve the charge transfer efficiency.\cite{ZhengKaibo2018Icae}
Moreover, recently experiments found that the introduction of S-S bond,\cite{RenHui2020EasR} halogen bond\cite{TremblayMarie} and $\rm{\pi}-\rm{\pi}$ stack\cite{GaoYao2019Meoo} into the spacer cation layer can enhance the intermolecular interactions which reduce the vdW gap.
These different organic spacer layers also cause changes for the electronic structure of inorganic layers near the band gap and affect the carrier transport process.\cite{StrausDanielB2019LCIE,GaoYao2019Meoo}
For instance, the introduction of aromatic cations with conjugated double bonds or thiophene rings with strongly polarized S atoms in the organic spacer cations can improve the energy alignment with the inorganic layer and the out of plane conductivity.\cite{NiChuyi2020TCIt,GaoYanbo2022HPLD}
However, the selection of different organic spacer cations is based on chemical intuitions and is mostly qualitative without a theoretical guidance.
More importantly, the fundamental mechanism of excited carriers transport has not yet been determined.
Particularly, in the experiment, owing to its ultra-fast property, the direct measurement of transport dynamics is challenging and many questions are unclear.
For example, what is the time scale of the carrier transfer, what is the role of the excitonic effect, and how do the various spacer molecules influence the carrier transfer at the atomic level.
To answer these questions, it is becoming extraordinary important to apply the first-principle calculations to quantify and analyze the carriers transport mechanism.

Non-adiabatic molecular dynamics (NAMD) simulation has becoming a powerful method to study the dynamics of excited carriers.\cite{RenJunfeng2013Nmds,KangJun2019Nmdw, ZhengFan2019UHCI, ZHAO2022}
In particular, the surface hopping method has been demonstrated to be an effective way to simulate the carrier dynamics directly.\cite{TULLYJ.C1990Mdwe}
The decoherence-induced surface hopping (DISH) method by realizing the detailed balance and the decoherence effect through the electron wavefunction collapsing, allows a real-time description of carrier transport and ionic movements.\cite{JaegerHeatherM2012Dsh,AkimovAlexeyV2013PECD,WangBipeng2022INMD}
Moreover, in most solid-state systems where the wavefunctions are usually delocalized, classical path approximation (CPA) can be applied by ignoring the ``feedback" of the excited carrier to the ionic motions, which can reduce the computation cost significantly.
In this case, the ground state ionic trajectory is used throughout the DISH simulation, and the NAMD becomes a post-processing of the ground state \textit{ab initio} MD.
This type of calculation allows for a direct simulation of a large system ($>$ 300 atoms) for a relatively long period ($\geq$ 1 picosecond), where the electron-phonon interaction is included automatically.
Thus, based on CPA, it is possible to explore various dynamical processes in a complex system, such as charge recombination\cite{LongRun2016UtEo,ZhangZhaosheng2019EDaS,ChuWeibin2020SLaD}, carrier transfer\cite{WangYing2021Nmda,ZhengQijing2017PUCT,ZhangLin2023CTDo}, carrier cooling \cite{NieZhaogang2014UCTa,Chu2022,ZhouZhaobo2022CoHC} etc.
In this work, by using the first-principle DISH implemented with CPA, we have studied the ultra-fast carrier transport across the spacer cation layer between single-inorganic-layer (\textit{n} = 1) and bi-inorganic-layer (\textit{n} = 2) 2D HOIPs.
By exploring different spacer cations, we find that the carrier transfer has a sub-picosecond time scale, and it depends on the spacer cation size significantly.
Different phases such as DJ and RP will also cause substantially different transfer time-scales.
Meanwhile, we have developed a new NAMD method to include the electron-hole Coulomb interaction, which could be significant for low dimensional materials.
Based on our simulation, we propose the appropriate spacer cation for fast carrier transfer.
We believe the first-principle theoretical analysis in this work will help to guide the experiments to improve the performance of 2D HOIP devices.

\begin{center}
  \includegraphics[scale=0.5]{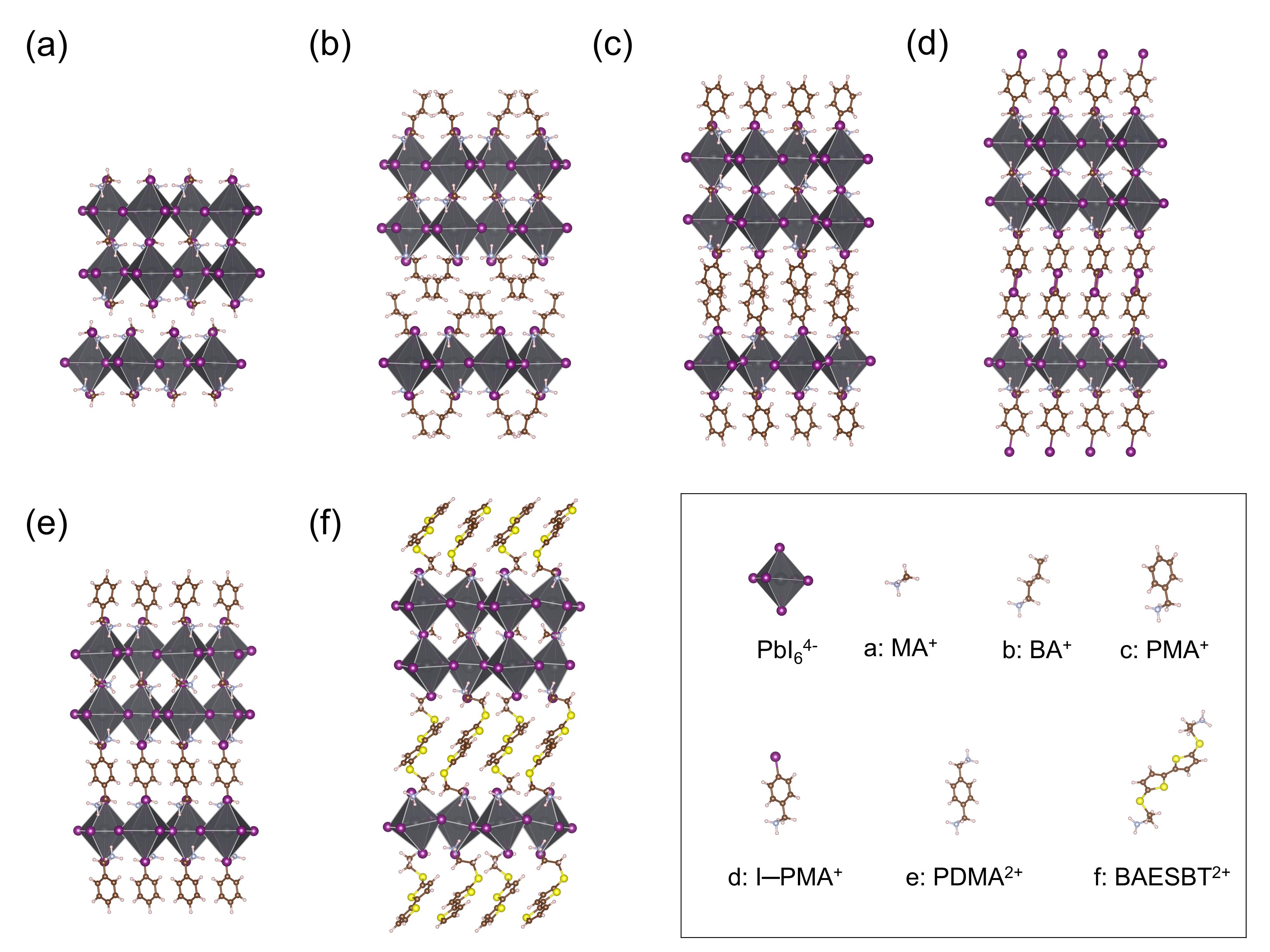}
\end{center}
Figure 1. (a-f) Equilibrium molecular structure of the $\rm{(MA)_{2}PbI_{4}}$/$\rm{(MA)_{3}Pb_{2}I_{7}}$, $\rm{(BA)_{2}PbI_{4}}$/ (BA$\rm{)_{2}(MA)Pb_{2}I_{7}}$, $\rm{(PMA)_{2}PbI_{4}}$/$\rm{(PMA)_{2}(MA)Pb_{2}I_{7}}$, (I-$\rm{PMA)_{2}PbI_{4}}$/(I-$\rm{PMA)_{2}(MA)Pb_{2}I_{7}}$, $\rm{(PDMA)PbI_{4}}$/$\rm{(PDMA)(MA)Pb_{2}I_{7}}$, $\rm{(BAESBT)PbI_{4}}$/$\rm{(BAESBT)(MA)Pb_{2}I_{7}}$ heterostructures. The insert at the bottom right shows the molecular structures of the inorganic ions and organic spacer cations of each structure. Grey atom: Pb, magenta atom: I, yellow atom: S,  brown atom: C, light blue atom: N, light pink atom: H.

Here, we build six types of 2D HOIP heterostructures with different spacer cations including DJ and RP phases; the spacer cation layer separates the two inorganic 2D HOIP with \textit{n} = 1 and \textit{n} = 2, respectively.
The equilibrium molecular structures of these six 2D HOIP heterostructures are shown in the Figure 1 with the following organic spacer cations: $\rm{MA^{+}}$ (methylammonium), $\rm{BA^{+}}$ (n-butylammonium),\cite{JungMi-Hee2021Eotp} $\rm{PMA^{+}}$ (phenylmethylammonium),\cite{JungMi-Hee2021Eotp} I-$\rm{PMA^{+}}$ (I-phenylmethylammonium),\cite{TremblayMarie} $\rm{PDMA^{2+}}$ (p-phenylenediamine),\cite{UmmadisinguAmita2022MSSo} and $\rm{BAESBT^{2+}}$ (5, 5'-bis(ammoniumethylsulfanyl)-2, 2'-bithiophene),\cite{ZhuXu-Hui2003EoMv} respectively.
The inorganic layers of these perovskite heterostructures are composed of $\rm{Pb^{2+}}$ and $\rm{I^{-}}$.
The spacer cation layer widths by measuring the I-I distances between the \textit{n} = 1 layer and \textit{n} = 2 layer of the six structures are 3\textup{~\AA}, 7\textup{~\AA}, 7.5\textup{~\AA}, 10.5\textup{~\AA}, 6\textup{~\AA}, and 10\textup{~\AA}, respectively.
Compared to the system with $\rm{MA^{+}}$ as the spacer cation, the other systems possess larger spacer distance:
$\rm{BA^{+}}$ has a long carbon chain, $\rm{PMA^{+}}$ introduces the benzene rings to form the $\rm{\pi - \pi}$ stack, while I-$\rm{PMA^{+}}$ replaces the H ion on the benzene ring with an iodine ion.
The $\rm{PDMA^{2+}}$ and $\rm{BAESBT^{2+}}$ are both bivalent cations containing the $\rm{\pi}-$conjugated group, forming the DJ phase 2D HOIPs.
First-principles DFT calculations are performed with the plane-wave package PWmat.\cite{JiaWeile2013Taoa,JiaWeile2013Fpwd}
The norm-conserving pseudopotentials are used with Perdew-Burke-Ernzerh of exchange-correlation functionals.\cite{PerdewJ.P.1998PBaE,HamannDR2013OnVp}
An energy cutoff of 50 Ryd is used to converge the charge density.
The empirical DFT-D3 method is used to capture the vdW interaction.\cite{GrimmeStefan2010Acaa}
The integration in the Brillouin zone is done with a $2 \times 2 \times 1$ k-point grid for all unit cells of the six 2D HOIP heterostructures. 
Then, the \textit{ab initio} ground state MDs are carried out using the $2\times2\times1$ supercells of the $\rm{MA_{2}PbI_{4}}$/$\rm{MA_{3}Pb_{2}I_{7}}$ heterostructure (hereinafter referred to as the MA system, and other heterostructures are similarly abbreviated), $\rm{BA}$ system, $\rm{PMA}$ system, I-PMA system, $\rm{PDMA}$ system, and $\rm{BAESBT}$ system.
They contain 432, 720, 752, 752, 656, and 880 atoms (Figure 1), respectively.
The MD calculations are performed with a single $\rm{\Gamma}$ k-grid at 300K with a time step $\Delta t$ = 1 fs.
During the \textit{ab initio} MD, eigen energies ($\epsilon_{i}(t)$), eigen wave functions ($\phi_{i}(\rm{\textbf{r}}$,$t)$), and the overlap between two adjacent eigen states $\left<\phi_{i}(t) \middle| \phi_{j}(t+\Delta t) \right>$ are stored for the post-processing NAMD calculation. 
See details of DISH implementation in SI, section S1.

\begin{center}
  \includegraphics[scale=0.1]{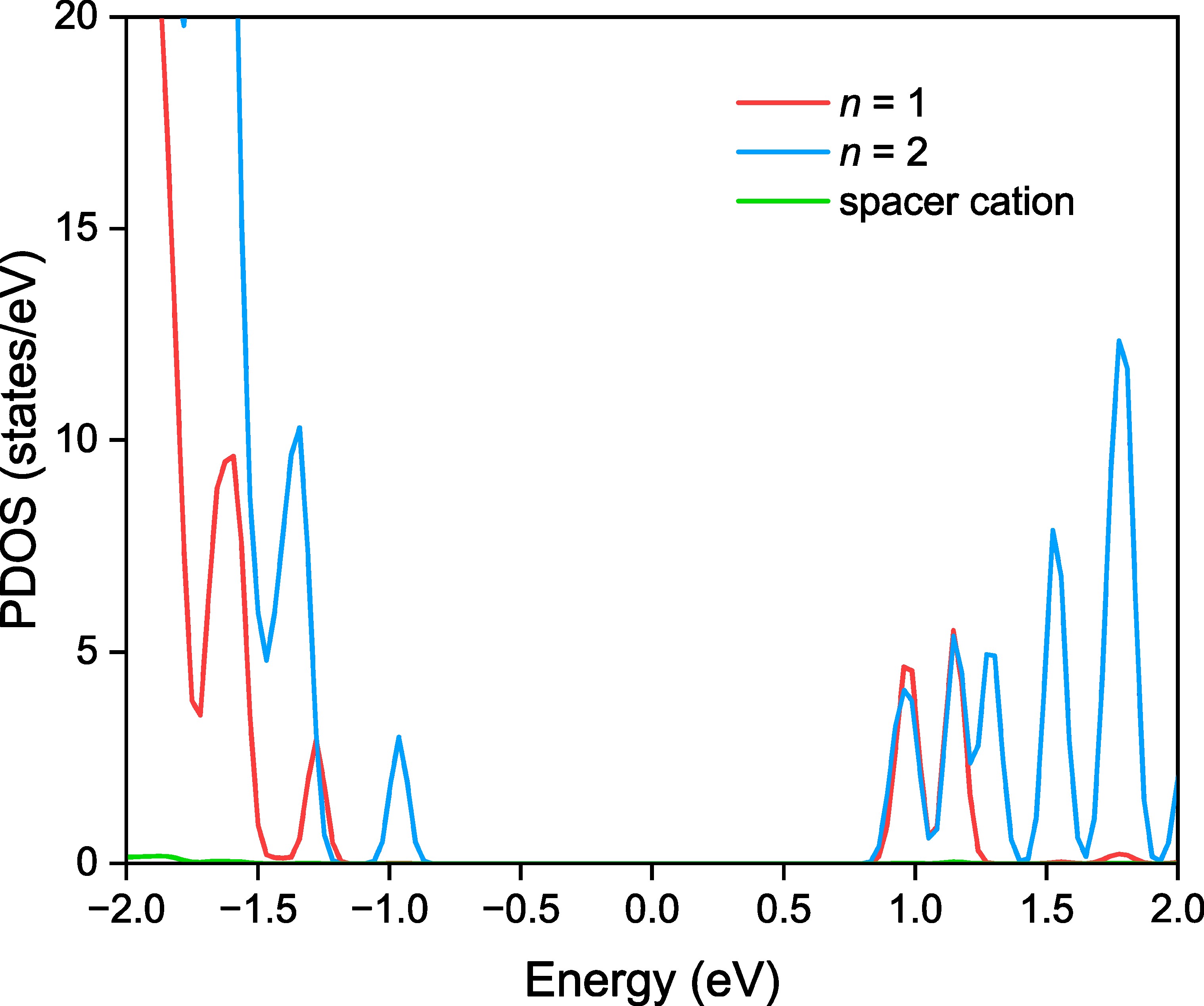}
\end{center}
Figure 2. Partial density of states (PDOS) of the $\rm{(MA)_{2}PbI_{4}}$/$\rm{(MA)_{3}Pb_{2}I_{7}}$ equilibrium heterostructure projected to \textit{n} = 1 layer, \textit{n} = 2 layer and spacer layer at 0K. 0 of x-axis is set as the Fermi level.

The partial density of states (PDOS) of the MA system at 0K, as displayed in Figure 2, shows that the molecular orbitals of $\rm{MA^{+}}$ are far away from the inorganic layer conduction band minimum (CBM) and valence band maximum (VBM), which is consistent to previous literature.\cite{ZhangLinghai2020BAiT}
The band offsets between the VBMs of \textit{n} = 1 layer and \textit{n} = 2 layer and between the CBMs of \textit{n} = 1 layer and \textit{n} = 2 layer are 0.31 eV and $-$0.05 eV, respectively, forming the type-$\rm{\uppercase\expandafter{\romannumeral1}}$ band alignment.
As aforementioned, in 2D HOIP heterostructures, the electrons and holes are likely to transfer from small \textit{n} layers to large \textit{n} layers by crossing the spacer cation layer.
Here, this process can be simulated directly in real-time using the NAMD method. By manually placing the electron at the CBM of \textit{n} = 1 layer and the hole at the VBM of \textit{n} = 1 layer, it is possible to track how these excited carriers transfer to the \textit{n} = 2 layer.
Charge distributions of the \textit{n} = 1 layer VBM and CBM are shown in Figure S1 which are the initial distributions of electron and hole charge in the NAMD simulation.

\begin{center}
\includegraphics[scale=0.12]{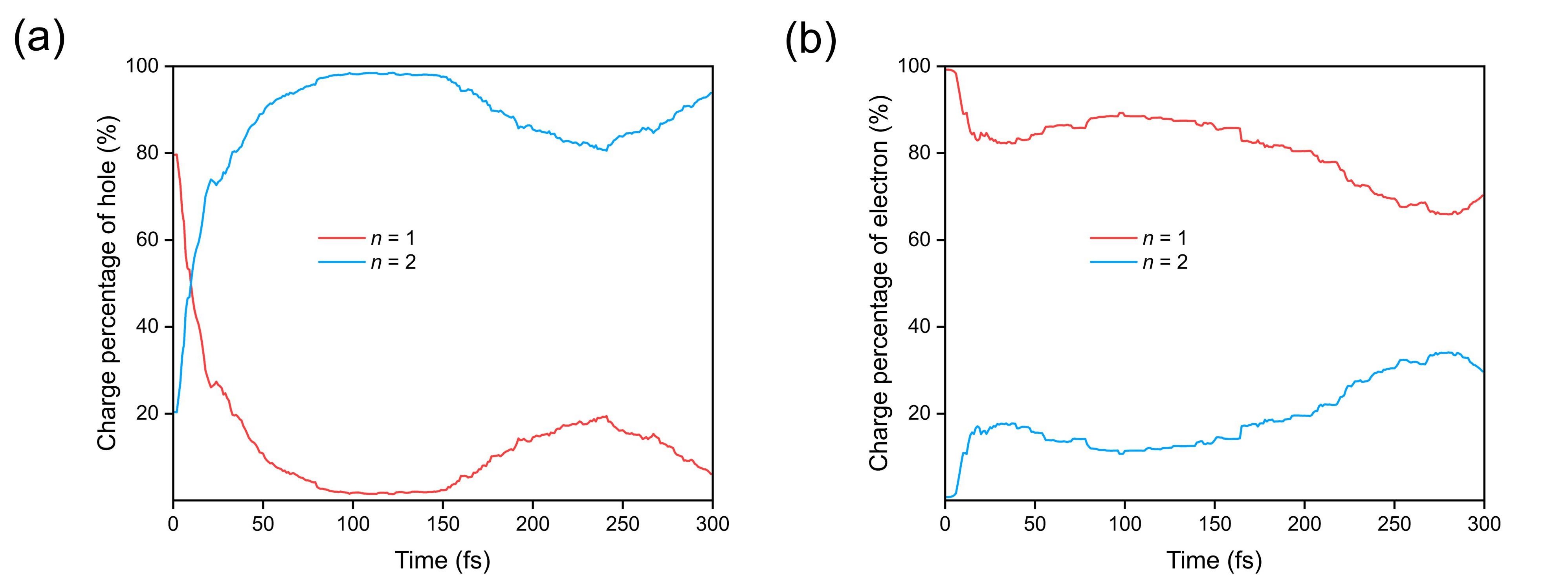} 
\end{center}
Figure 3. (a) Hole distribution in \textit{n} = 1 layer and \textit{n} = 2 layer of the MA system. (b) Electron distribution in \textit{n} = 1 layer and \textit{n} = 2 layer of the MA system.

Shown in Figure 3 is the charge distribution of the hole and the electron in \textit{n} = 1 layer and \textit{n} = 2 layer.
Owing to the stochastic nature of DISH algorithm, a large number of independent trajectories are performed. 
The charge distribution here is averaged over these trajectories.
Our calculations show that the carrier transfer between 2D HOIP heterostructures is indeed an ultra-fast process which is on the sub-picosecond time scale.
Here, the hole tends to present a much faster transfer rate compared to the electron.
This is owing to the fact that the electron is transferring from CBM of \textit{n} = 1 layer to CBM of \textit{n} = 2 layer, where these two states are mainly located on the Pb-atoms of the \textit{n} = 1 and \textit{n} = 2 inorganic layers separated by the spacer cations and the I-atoms.
However, for the hole transfer, the VBM of \textit{n} = 1 layer and VBM of \textit{n} = 2 layer are mostly hybridized between Pb \textit{s} and I \textit{p} orbitals.
In this case, their couplings between VBM of \textit{n} = 1 layer and VBM of \textit{n} = 2 layer are stronger due to the closer distance compared to the electron case.
In addition, the fluctuation of the charge occupation (electron and hole) is interesting to explore.
Particularly, for the hole, after its transfer from \textit{n} = 1 layer to \textit{n} = 2 layer (time $>$ 100 fs), part of the hole returns to \textit{n} = 1 layer around time 250 fs.
Meanwhile, for the electron, its transfer shows a complex function of time.
We find that such fluctuation is strongly related to the adiabatic-state fluctuation at a finite temperature.
As shown in Figure S3, the eigen energies and the adiabatic states (belong to \textit{n} = 1 layer or \textit{n} = 2 layer) are illustrated as a function of time. 
The back transfer of the hole around 250 fs is caused by the major occupation of states of \textit{n} = 1 layer in the valence bands region.
For the electron, due to the fluctuation of the adiabatic states, the system becomes type-$\rm{\uppercase\expandafter{\romannumeral2}}$ interface up to 150 fs.
During this time period, the electron transfer is suppressed since the CBM of \textit{n} = 1 layer is lower than the CBM of \textit{n} = 2 layer.
The electron transfer rate is accelerated after 150 fs when the CBMs of \textit{n} = 1 layer and \textit{n} = 2 layer have similar energies.
Here, our finding is consistent to the previous work discussing the band fluctuation at a finite temperature.\cite{ZhangLinghai2020BAiT}
In the realistic cases, where higher-\textit{n} 2D HOIPs will exist, the band offset between different \textit{n}-layer HOIPs could surpass the fluctuation of adiabatic state at a finite temperature.
We believe that in these cases the aforementioned hole back-transfer and the complex electron transfer pattern will not play a role.

Since most of the NAMD simulations rely on the ground state Hamiltonian, the single-particle picture is used throughout the simulation and the excitonic effect is mostly ignored.
This treatment is generally fine for bulk materials with strongly screened electron-hole interaction,
but such interaction becomes critical in low-dimensional materials.\cite{ShiRan2020Clco}
Typically, the large energy barrier of the organic spacer restricts the excitons to the 2D HOIPs QW planes, which makes it difficult for carriers to transport between inorganic layers.
For instance, Prezhdo et al found nonradiative electron-hole recombination constitutes a major pathway for charge and energy losses in 2D HOIPs.\cite{doi:10.1021/acs.nanolett.8b00035}
Using the versatile optical spectroscopy measurements, Xiong et al elucidated that the the exciton fine-structure splitting in 2D HOIPs is attributed to this enhanced electron-hole exchange.\cite{DoT.ThuHa2020BEFi}
Here, we approximate the electron-hole interaction as purely the Coulomb interaction, which is also the dominant interaction of electrons and holes in nano materials and this interaction is added to the NAMD Hamiltonian during the dynamics.
In practice, the Coulomb potential by the hole applied to the electron is taken into account by evolving the time-dependent electronic Hamiltonian as:
\begin{align}
  H_{\rm{el}}^{'}(\textbf{r}_{\rm{el}}, \textbf{R}(t)) & = H_{\rm{el}}(\textbf{r}_{\rm{el}},\textbf{R}(t)) + V_{\rm{el}}(\rho_{\rm{hole}}(t)),
\end{align}
where $H_{\rm{el}}$ is the single-particle electronic Hamiltonian in the evolution of the time-dependent Schr$\ddot{\rm{o}}$dinger equation (TDSE) for the electron, $\rm{\textbf{R}}(t)$ is the time-dependent trajectory, $V_{\rm{el}} (\rho_{\rm{hole}}$ $(t))$ is the \textit{screened} potential by the hole acting on the electron, and $H_{\rm{el}}^{'}$ is the Hamiltonian including the electron-hole interaction.
Meanwhile, the potential ($V_{\rm{hole}}$) by the electron charge density ($\rho_{\rm{el}}(t)$) is computed simultaneously and is applied to the hole Hamiltonian.
Here, the evaluation of the \textit{screened} Coulomb potential $V_{\rm{el}}$ (or $V_{\rm{hole}}$) is challenging, since the screening effect can not be approximated by a single homogeneous function in this 2D setup.
To capture this effect efficiently, the time-dependent charge density of the electron $\rho_{\rm{el}}(t)$ (or hole) is decomposed into a set of so-called ``trial charge density" with the time-dependent coefficient.
The total time-dependent screened potential can be computed as the linear combination of the ``trial charge potentials" computed based on the initial ``trial charge densities" with the appropriate weights.
We apply this method for the time-dependent charge density of electron and hole simultaneously to include their Coulomb interactions.
More details can be obtained in SI.

By adding the electron-hole interaction into the simulation, as the Figure S10 shows, for the initial structure with occupying the CBM and VBM of the \textit{n} = 1 layer, respectively, the excitonic effect increases the VBM of \textit{n} = 2 layer by 0.19 eV and increases the CBM of \textit{n} = 2 layer by 0.09 eV, changing the band alignment into the type-$\rm{\uppercase\expandafter{\romannumeral2}}$.
By implementing the time-dependent electron-hole interaction in the whole NAMD process, we find that the hole and electron transport from \textit{n} = 1 layer to \textit{n} = 2 layer across organic spacer layer is not strongly affected by the electron-hole interaction (Figure S11).
This is because the excitonic effect is relatively small.
Although we are studying a 2D system, the strong dielectric screening of HOIPs and the relatively large thickness do not make a strong excitonic contribution to the carrier dynamics.
Meanwhile, owing to the small eigen-energy changes by the excitonic effect, the eigen-energy fluctuations still play a role.
Thus, for the following calculations, the excitonic effect is not added.

\begin{center}
  \includegraphics[scale=0.5]{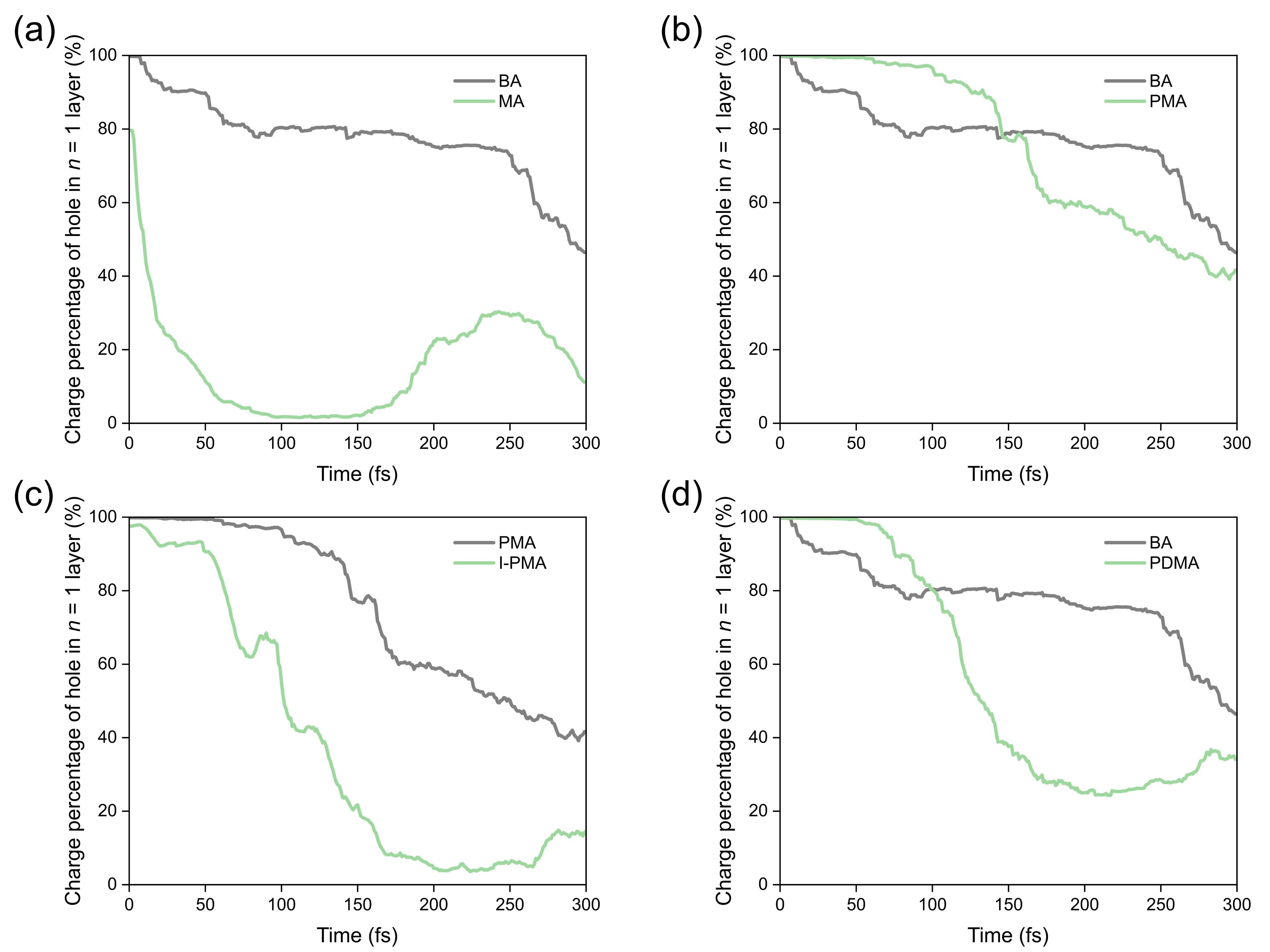}
\end{center}
Figure 4. Hole distribution in \textit{n} = 1 layer of (a) the MA system vs. the BA system, (b) the BA system vs. the PMA system, (c) the PMA system vs. the I-PMA system, and (d) the BA system vs. the PDMA system.

To explore the influence of different spacer cations to the carrier transport process, four more HOIP systems are studied.
Figure 4 shows the time evolution of hole distribution in \textit{n} = 1 layer of these systems, and more details are shown in Figure S13-16.
As Figure 4a and Figure S13b-c show, the carrier transfer rate in the $\rm{BA}$ system is significantly slower than that in the $\rm{MA}$ system.
This result is consistent with the carrier transfer rate in the $\rm{MA}$ system with an increased spacer-cation-layer width of 4 \textup{~\AA} (as shown in Figure S12).
In this case, the distance is a critical factor for the carrier transfer rate, and a smaller-spacer cation (e.g. $\rm{MA^{+}}$) is preferred.
However, the aliphatic spacer cations (e.g. the MA system) lack interactions between these molecules, which may lead to the penetration of water molecules and air entering the perovskite layer and destabilize the structure.\cite{Juarez-PerezEmilioJ2016TdoC,https://doi.org/10.1002/aenm.201500477}
Especially, $\rm{MA^{+}}$ is demonstrated to volatilize and escape from the perovskite crystal lattice easily, which would decrease the device thermal and light stability.
Thus, although the $\rm{MA}$ system has a ultra-fast carrier transfer rate, it is not the best choice for the efficient carrier transport.
Compared to the aliphatic series, the aromatic spacer cations contain the $\rm{\pi-\pi}$ stack enhance the stability of the perovskite layer and are more suitable for high performance devices.\cite{QianChong-Xin2022Fo2p,MuscarellaLoretaA2022RPMo} 
For instance, the $\rm{PMA}$ system has been demonstrated to exhibit excellent stability.
The spacer-cation-layer width of the $\rm{PMA}$ system is 7.5 \textup{~\AA}, which is almost the same to the $\rm{BA}$ system.
Figure 4b also shows the similar carrier transfer rates of these two systems which further confirms that distance has a decisive influence on carrier transfer rate.

It is reported that the addition of halogen atoms to the organic spacer layer can suppress the ion migration and improve the structural stability of 2D HOIPs, which is a promising research direction.\cite{FuXinliang2023MPwH}
In addition to the stability enhancement, our calculations reveal that the addition of halogen atoms into the spacer cation layer, acting as a carrier transport bridge, will also benefit the carrier transport process.
Figure S15a shows the PDOS of the I-PMA system.
Different from the aforementioned 2D HOIP systems, the molecular orbitals of I-$\rm{PMA^{+}}$ are close to the VBM of inorganic layers and the adiabatic states also shows the charge distribution in the spacer cation layer along the trajectory.
Compared to the $\rm{PMA}$ system, as shown in Figure 4c, the I-PMA system has a much faster hole transfer rate.
Figure S15c shows that when the charge distribution in the \textit{n} = 1 layer starts to decrease around 110 fs, the charge distribution in the spacer cation layer starts to increase.
After 110 fs, with the charge distribution in the spacer cation layer decreasing, the charge continues to transfer to the \textit{n} = 2 layer.
Here, we have shown in real-time that the spacer cation layer acts as a bridge to assist the hole transport.
However, owing to the larger Pb-Pb distance separated by the large spacer cation layer and I atoms, the electron hardly transport in the I-PMA system.

Based on the existing research findings, DJ phase 2D HOIPs may be competitive candidates to achieve a high carrier transfer rate.\cite{JiangXiaoqing2020D2ps,GhoshDibyajyoti2020Ccdi,ZhangBobo2023ECSb}
On the one hand, the interaction between different inorganic layers can be significantly enhanced by the shorter spacer cation layer width.
On the other hand, the rigid structure of DJ phase 2D HOIPs will result in both low defect density and low ion migration.\cite{ZhangYunxia20202PSC}
As shown in Figure 4d, the DJ phase $\rm{PDMA}$ system shows a faster hole transfer rate than the $\rm{BA}$ system, although their spacer-cation-layer widths are similar.
Moreover, the electron transfer rate, as shown in Figure S16d, is also slightly enhanced compared to the $\rm{PMA}$ and I-PMA systems.
However, the stability of DJ phase 2D HOIPs is highly dependent on the rigidity of organic cations.\cite{LIU20231016}

\begin{center}
\includegraphics[scale=0.5]{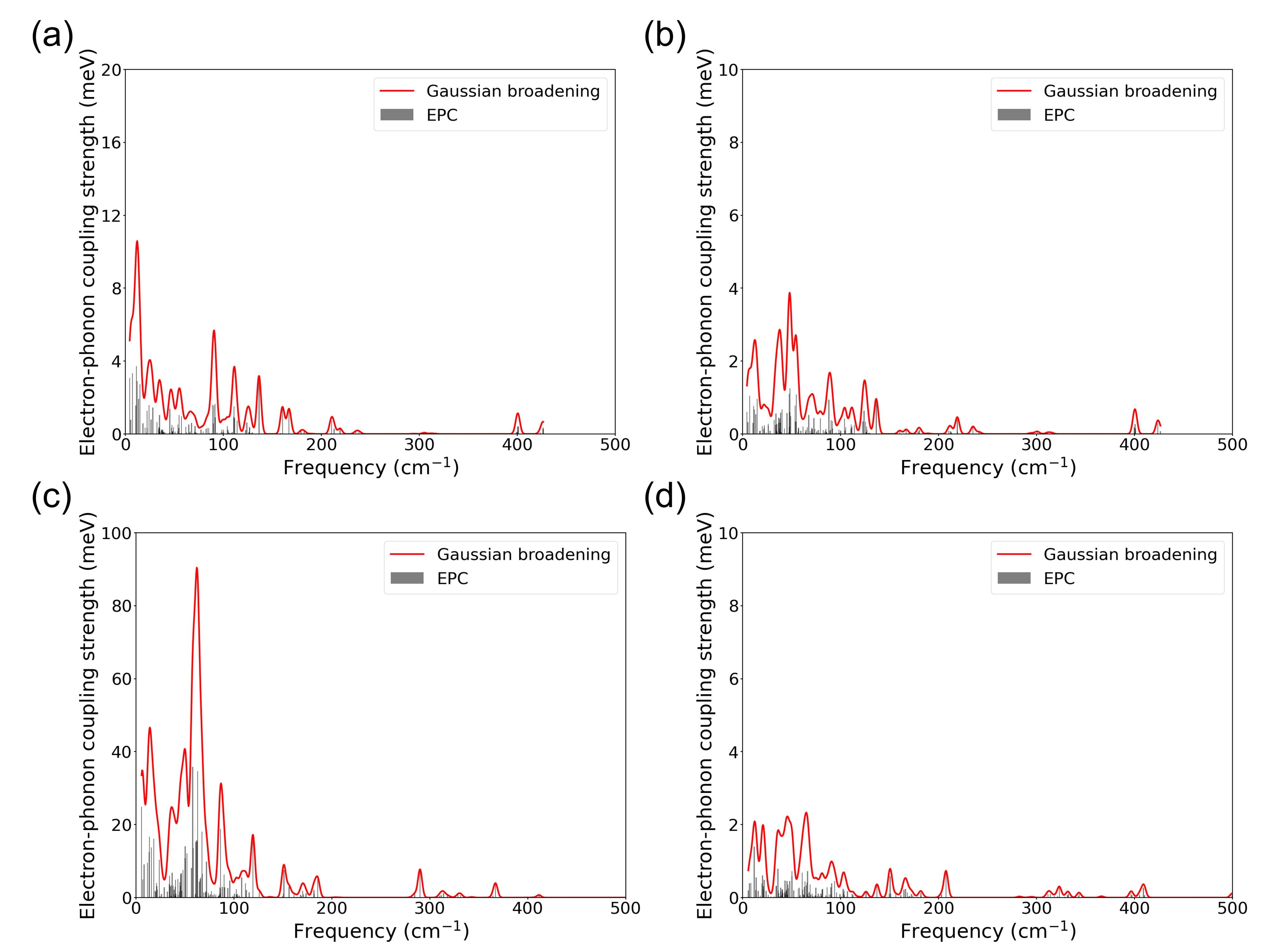}
\end{center}
Figure 5. (a)The EPC strengths between the VBM of \textit{n} = 1 layer and the VBM of \textit{n} = 2 layer of the RP (the $\rm{BA}$ system). (b)The EPC strengths between the CBM of \textit{n} = 1 layer and the VBM of \textit{n} = 2 layer of the RP (the $\rm{BA}$ system). (c)The EPC strengths between the VBM of \textit{n} = 1 layer and the VBM of \textit{n} = 2 layer of the DJ (the $\rm{PDMA}$ system). (d)The EPC strengths between the CBM of \textit{n} = 1 layer and the VBM of \textit{n} = 2 layer of the DJ (the $\rm{PDMA}$ system).

Here, it is interesting to reveal why the DJ phase 2D HOIPs are more beneficial to carrier transport compared to the RP phase.
This can be explained by computing the electron-phonon coupling (EPC) between the VBMs (and CBMs) of \textit{n} = 1 layer and \textit{n} = 2 layer, which is directly related to the aforementioned non-adiabatic coupling.\cite{FanZheng2022Amkn}
The EPC matrix element is given by:\cite{GiustinoFeliciano2017Eiff}
\begin{align}
  g_{mn\nu} & = \left< u_{m}\middle| \Delta_{\nu} v^{KS}\middle | u_{n}\right>_{\rm{sc}},
\end{align}
here, the subscript ``sc" means that the integral is carried out within the supercell which is the simulation cell used in this work.
The $u_{m}$ and $u_{n}$ are the periodic part of the Kohn-Sham electron wave functions of band $m$ and $n$, $\Delta_{\nu}v^{\rm{KS}}$ is the phonon mode $\nu$ induced variation of the Kohn-Sham self-consistent potential, which can be obtained by a standard DFT SCF.
More details can be obtained from SI.
Here, we compute the EPC strengths between \textit{n} = 1 and \textit{n} = 2 VBMs (and CBMs) for $\rm{BA}$ (RP phase) and $\rm{PDMA}$ (DJ phase) systems which have similar I-I distances.
Figure 5 shows the EPC strengths of the two HOIP systems in the frequency range of 0-500 $\rm{cm^{-1}}$.
The EPC strengths between the VBMs of \textit{n} = 1 layer and \textit{n} = 2 layer of the PDMA system are much larger than that of the BA system, while the EPC strengths between the CBMs are similar in the two systems.
This result is consistent to the faster hole transfer rate in the PDMA system than that in the BA system, and the similar electron transfer rate for both systems.

\begin{center}
  \includegraphics[scale=0.5]{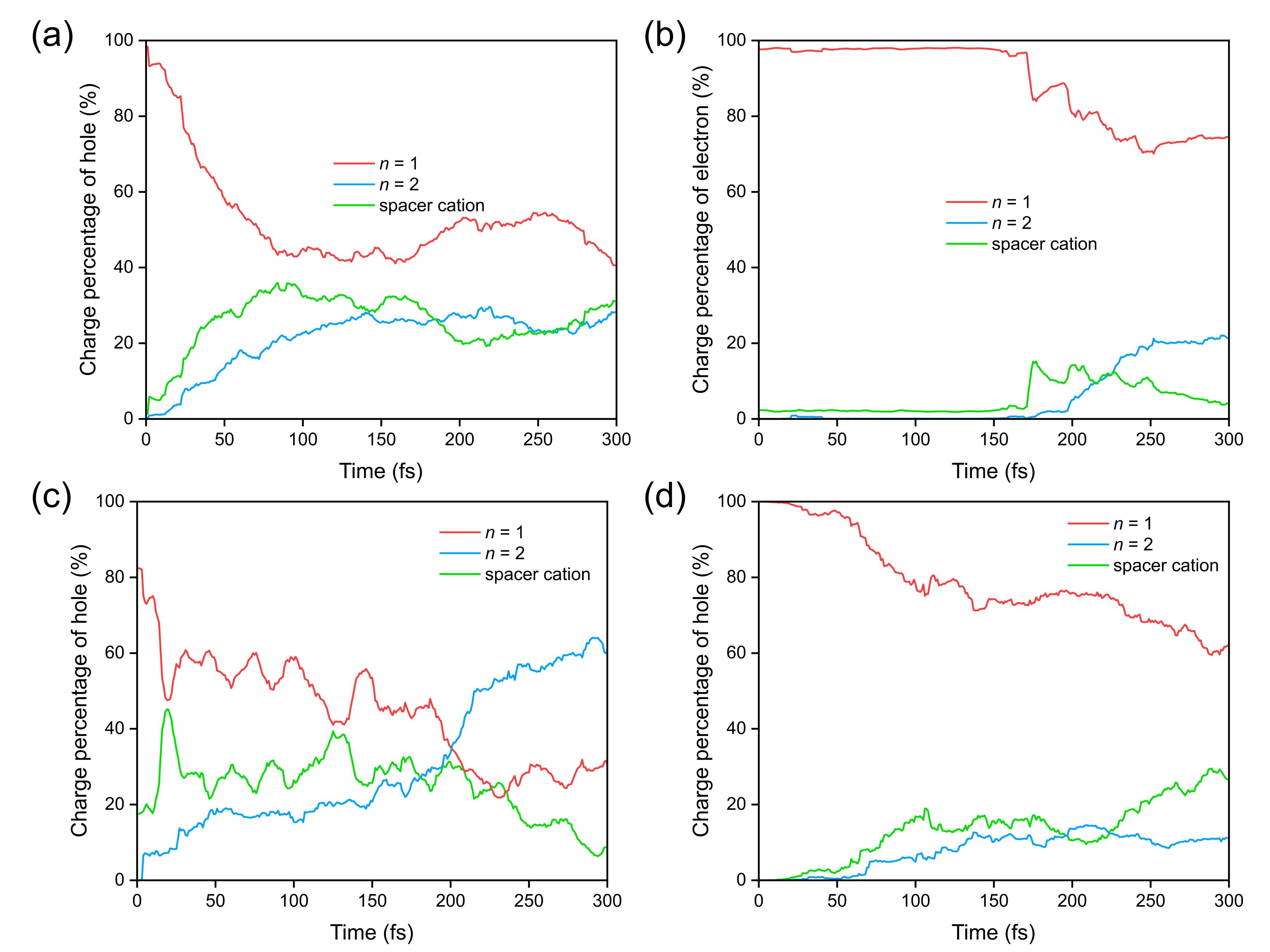}
\end{center}
Figure 6. (a) Hole distribution in \textit{n} = 1 layer, \textit{n} = 2 layer, and spacer layer of the BAESBT system. (b) Electron distribution in \textit{n} = 1 layer, \textit{n} = 2 layer, and spacer layer of the BAESBT system. (c) Hole distribution in \textit{n} = 1 layer, \textit{n} = 2 layer, and spacer layer of the $\rm{CF}_{3}$-BAESBT system. (d) Electron distribution in \textit{n} = 1 layer, \textit{n} = 2 layer, and spacer layer of the $\rm{CF}_{3}$-BAESBT system.

Based on our calculations, in order to obtain 2D HOIP with an efficient carrier transfer rate, a stable DJ phase structure combined with a bridge atom to assist both hole and electron transfers is preferred.
Thiophene-based molecule, which improves the band edge alignment and enhances the charge transport of perovskites, can be one of the appropriate bridge groups.\cite{NiChuyi2020TCIt}
However, the single thiophene ring with destabilized aromatic structure forms a RP phase 2D HOIP.
It has been found that the introduction of bithiophene ring leads to the formation of a stable DJ phase 2D HOIP.\cite{WangKang2021LOHQ,DongYixin2023OIbt}
On the one hand, the $\rm{\pi}$-conjugated bithiophene ring molecules are hydrophobic, effectively blocking the entry of air and water molecules. 
On the other hand, the interactions between the $\rm{\pi}$-conjugated molecules stabilize the perovskite structure.\cite{GaoYao2019HSLP}
Zhu et al successfully synthesize a di-ammonium cation of $\rm{2,\ 2^{'}}$-Bithiophene derivative ($\rm{(BAESBT)PbI_{4}}$) in DJ phase.\cite{ZhuXu-Hui2003EoMv}
The S atoms connect the bithiophene rings and $\rm{NH_{3}^{+}}$ which further increase the molecule highest occupied molecular orbital (HOMO) and enhance the hole transfer.\cite{SunQingde2019DCDP}
Here, the $\rm{BAESBT}$ system is built and the computed PDOS is shown in Figure S18a.
The molecular orbitals of $\rm{BAESBT^{2+}}$ are both close to the VBM and CBM of inorganic layers, and we believe this type of spacer molecules will benefit both electron and hole transfer.
This is indeed what we have found.
As shown in Figure 5a-b, although the spacer cation width of BAESBT system is much larger than that of BA system, the electron and hole transfers are greatly improved.
However, we noticed that before 150 fs, no electron can transfer.
Meanwhile, although the electron transfer is elevated, it is still not comparable to the hole transfer.
This is owing to the fact that the spacer molecular orbitals are still too high for the CBM of the inorganic layers.
Here, we propose to introduce the electron-absorbing groups, such as $\rm{CF_{3}^{+}}$, on the bithiophene ring, thereby reducing the energy level of organic molecules (Figure S20).
As shown in Figure 6c-d, for the $\rm{CF_{3}}$-BAESBT system, the major part of the hole has been transferred to \textit{n} = 2 layer.
For electrons, the transfer rate is enhanced greatly and the electrons start to transport to \textit{n} = 2 layer at the beginning of the trajectory.
We also analyze the electron-phonon coupling for this system.
Figure S21 shows the EPC strengths of the VBMs (and CBMS) of \textit{n} = 1 layer and \textit{n} = 2 layer of the BAESBT system.
Different from the PDMA or BA systems, the EPC strengths of the BAESBT system have shown an enhanced coupling for the conduction bands.

In conclusion, 2D HOIPs with different sizes and different types of organic spacer cations were simulated by the real-time first-principle NAMD method, and the carrier transport process of each structure was analyzed quantitatively. 
We found that the carriers transport across the spacer cation layer between the $\rm{PbI_{4}^{2-}}$ inorganic layers is an ultra-fast process with a sub-picosecond time scale.
We have developed a method to include the electron-hole interaction into NAMD.
By comparing different spacer cations, we found the size of the organic spacer layer largely determines the carrier transport rate.
Moreover, by introducing the bridge atoms (e.g. iodine atoms) or group (e.g. bithiophene ring) in the spacer cation layer, our calculations have demonstrated the significant enhancement of the carrier transfer through the bridges.
Additionally, the DJ phase was found to better assist the carrier transport across the spacer layer.
This is further investigated by comparing the electron-phonon coupling strengths.
By combing the above two advantages, we propose the $\rm{BAESBT^{2+}}$ cation including the $\rm{CF_{3}}$-$\rm{BAESBT}^{2+}$ as the spacer cation layers.
Our calculations have demonstrated that this type of spacer molecule provides an efficient transport for both electrons and holes.
We hope our finding can guide the experiments to select suitable organic spacer cations towards high performance optoelectronic devices.

\textbf{Acknowledgements}

This work is supported by National Natural Science Foundation of China (No. 62305215) and the ShanghaiTech University. 
The computational support is provided by HPC Platform of ShanghaiTech University.



\end{document}